\documentclass[pra,twocolumn,superscriptaddress,aps,showpacs,amsmath,amssymb,nofootinbib]{revtex4}

\pdfoutput=1
\usepackage{color}
\usepackage{graphicx}
\usepackage{bm}
\newcommand{\ignore}[1]{}
\def\beq{\begin{equation}}
\def\eeq{\end{equation}}
\def\beqa{\begin{eqnarray}}
\def\eeqa{\end{eqnarray}}

\begin{document}
\title{Hybrid synchronization in coupled ultracold atomic gases}

\author{Haibo Qiu}
\affiliation{Departament d'Estructura i Constituents de la Mat\`{e}ria,\\
Universitat de Barcelona, 08028 Barcelona, Spain}
\affiliation{College of Science, Xi'an University of Posts and
Telecommunications, 710121 Xi'an, China}
\affiliation{Institut de Ci\`encies del Cosmos, 
Universitat de Barcelona, IEEC-UB, Mart\'i i Franqu\`es 1, E--08028
Barcelona, Spain}

\author{Roberta Zambrini} 
\affiliation{Instituto de F\'isica Interdisciplinar y Sistemas Complejos IFISC (CSIC-UIB),
Campus Universitat Illes Balears, E-07122 Palma de Mallorca, Spain}

\author{Artur Polls}
\affiliation{Departament d'Estructura i Constituents de la Mat\`{e}ria,\\
Universitat de Barcelona, 08028 Barcelona, Spain}
\affiliation{Institut de Ci\`encies del Cosmos, 
Universitat de Barcelona, IEEC-UB, Mart\'i i Franqu\`es 1, E--08028
Barcelona, Spain}

\author{Joan Martorell} 
\affiliation{Departament d'Estructura i Constituents de la Mat\`{e}ria,\\
Universitat de Barcelona, 08028 Barcelona, Spain}

\author{Bruno Juli\'a-D\'\i az}
\affiliation{Departament d'Estructura i Constituents de la Mat\`{e}ria,\\
Universitat de Barcelona, 08028 Barcelona, Spain}
\affiliation{Institut de Ci\`encies del Cosmos, 
Universitat de Barcelona, IEEC-UB, Mart\'i i Franqu\`es 1, E--08028
Barcelona, Spain}
\affiliation{ICFO-Institut de Ci\`encies Fot\`oniques, 
Parc Mediterrani de la Tecnologia, 08860 Barcelona, Spain}

\date{\today}
\begin{abstract}
We study the time evolution of two coupled many-body quantum systems 
one of which is assumed to be Bose condensed. Specifically, we consider 
two ultracold atomic clouds each populating two localized single-particle 
states, i.e. a two-component bosonic Josephson junction. 
The cold atom cloud can retain its coherence when coupled to the condensate
and displays synchronization with the latter, differing from usual entrainment.
We term this effect among the ultracold and the condensed clouds as 
{\it hybrid synchronization}. The onset of synchronization, which we 
observe in the evolution of average properties of both gases when 
increasing their coupling, is found to be related to the many-body 
properties of the quantum gas, e.g., condensed fraction, quantum fluctuations of the particle number differences. We discuss the effects of different initial preparations, and the influence of unequal particle numbers for the two clouds, and we explore the dependence on the initial quantum state, e.g. coherent state, squeezed state, and Fock state, finding essentially the same phenomenology in all cases.
\end{abstract}

\maketitle

\section{Introduction}

Synchronization has been described in physics, chemistry, 
biology and social behavior~\cite{Review,Review2,Review4}. It has been extensively studied in classical non-linear dynamical 
systems~\cite{Review3}, and chaotic ones~\cite{Chaos}. The same 
phenomena have been explored recently in quantum systems, e.g. 
opto-mechanical devices~\cite{OptMec},
damped  harmonic oscillators~\cite{Syn6,Zam2},  
driven \cite{Syn3} and purely dissipative 
spins~\cite{Syn4}, and non-linear optical cavities~\cite{Syn5}. 
Synchronization can refer to
the mutual effect between detuned but otherwise equivalent components adjusting their rhythms (spontaneous
synchronization) as, for instance in Refs.~\cite{Syn6,Zam2,Syn4}.
Otherwise, a slave system can be driven to follow the dynamics of an external source 
leading to entrainment or driven synchronization, as for instance in Refs.~\cite{OptMec,Syn3}. In quantum 
many-body physics connections between quantum entanglement and 
mutual synchronization have been discussed in continuous variable 
systems ~\cite{Syn6,Zam2,Syn7}.

Ultracold atomic gases are particularly relevant quantum many-body 
systems. Since the first experimental production of Bose-Einstein 
condensates (BEC's), they have evolved from being 
a theoretical curiosity to versatile systems potentially useful in 
a large number of fields~\cite{lewenstein-book}. Identifying the onset 
of synchronization in these systems and proposing ways in which such 
phenomena can be characterized both experimentally and theoretically 
is a significant step forward in our understanding of the dynamical 
evolution of coupled quantum many-body systems.

Among the most promising applications are those that stem from the 
macroscopic sizes of the condensates. BECs are fantastic candidates for high 
accuracy interferometric devices~\cite{cronin09,schaff15}. These 
devices rely on the high degree of coherence maintained by BECs. 

In recent years, experiments with bi-modal ultracold atomic gases have 
managed to produce entangled ensembles in which the interferometric 
capabilities can be largely 
enhanced~\cite{giova04,esteve08,gross10,riedel10,bohnet14}. These improved 
interferometric properties are directly related 
to the pseudo-spin squeezing which can be produced using 
several techniques~\cite{kita93,Bo97,julhei12}. One of the main sources of 
decoherence in BECs are atom-atom interactions which induce 
dephasing of the different Fock components~\cite{Lew96}. Several 
possibilities, notably the generation of squeezed states~\cite{Bo97}, 
have been proposed to increase the coherence times and improve 
the interferometric capabilities, e.g. of the recent Mach-Zehnder proposal~\cite{berrada13}. 

In this paper we describe how decoherence effects due to the 
atom-atom interaction can be largely suppressed if a quantum many-body 
system is coupled to a Bose-Einstein condensate. To be more specific, 
we consider two bosonic Josephson junctions, $a$ and $b$. Subsystem 
$b$ is taken to be a BEC at all times during the evolution~\cite{Smerzi97}, 
while subsystem $a$ is a standard bosonic Josephson junction, i.e. it may 
fragment during the evolution~\cite{Milburn97,jame05,bruno10}. 
The coupling between $a$ and $b$ is provided by the contact inter-species interaction.
Directly related to this improvement in the coherence times of subsystem $a$, 
is the onset of a hybrid synchronization between $a$ and $b$. 

This synchronization is called hybrid because the two coupled atomic samples 
are in different regimes, one being a BEC described within a classical 
approximation and the other being a cloud of cold atoms described with a fully 
quantum formalism. We notice that in the literature the term {\it hybrid synchronization} 
has also been used in other contexts to describe, for instance, synchronization 
between (classical) chaotic systems persisting despite the difference 
in some variables \cite{hybrid1} or to the co-existence of different kinds 
of synchronization in composed (classical) systems  like cascade-coupled 
lasers Ref.~\cite{hybrid2}.

The paper is organized as follows: In Sect.~\ref{sec:model} we introduce 
the model system, a two species two-site Bose-Hubbard model. Assuming that one 
of the species is condensed, we construct our mixed quantum-classical 
description. In Sec.~\ref{sec:dyn} the coupled dynamics is studied, 
particularly the onset of hybrid synchronization and its relation to 
the condensed fraction of subsystem $a$. Finally, in Sec.~\ref{sec:conc} we 
present our conclusions.

\section{Model}
\label{sec:model}

Let us consider two different kinds of bosons, $a$ and $b$, populating two 
single-particle states, $L$ and $R$. The main results described in this paper 
are essentially independent of the nature of the single-particle states. A way 
to produce them is by trapping ultracold atoms in a spatial double-well 
potential~\cite{GO07,esteve08}. A different possibility would be to populate 
two hyperfine states of the atom and couple them linearly as in Ref.~\cite{zib10}. 
A solid-state alternative could be provided by extending the experiments 
in Ref.~\cite{amo13} to two different kinds of exciton-polaritons. 

The interaction between the bosons is assumed to be described by a contact 
term with strength proportional to the $s-$wave scattering length. For 
simplicity the bosons are taken to be two different hyperfine states of 
the same atom, so that the masses of both species $a$ and $b$, is the same. 
With these simplifications, the many-body, Bose-Hubbard, Hamiltonian for 
the system reads,
\beqa
\hat{\cal H}= \hat{\cal H}_a + \hat{\cal H}_b + \hat{\cal H}_{ab}
\eeqa
where
\beqa
\hat{\cal H}_a&=& 
- J_a (\hat{a}^\dagger_L \hat{a}_R + \hat{a}_L \hat{a}^\dagger_R )  \nonumber\\&+&
{U_a\over2} ( \hat{n}_{a,L} (\hat{n}_{a,L}-1) + \hat{n}_{a,R} (\hat{n}_{a,R}-1) ), \nonumber\\
\hat{\cal H}_b&=& - J_b (\hat{b}^\dagger_L \hat{b}_R + \hat{b}_L \hat{b}^\dagger_R )  \nonumber\\&+&
{U_b\over2} \left( \hat{n}_{b,L} (\hat{n}_{b,L}-1)
+ \hat{n}_{b,R} (\hat{n}_{b,R}-1) \right), \nonumber\\
\hat{\cal H}_{ab} &=& 
{U_{ab}\over2}\left(\hat{n}_{a,L}- \hat{n}_{a,R})(\hat{n}_{b,L}-\hat{n}_{b,R} \right)  \,.
\eeqa
$\hat{a}_{L(R)}^{\dag }$$(\hat{a}_{L(R)})$ and $\hat{b}_{L(R)}^{\dag }$$(\hat{b}_{L(R)})$ 
are creation (annihilation) operators for the $L$ or $R$ modes of $a$ and $b$. 
The Hamiltonian includes tunneling terms, proportional to $J_{a(b)}$, which the in absence of any interaction induce periodic Rabi oscillations of the populations between the states. The contact interaction translates into terms with strength proportional to $U_a$, $U_b$, and $U_{ab}$ which gauge the $aa$, $bb$ and $ab$ 
contact interactions. The coupling between the two gases is solely due to the term 
proportional to $U_{ab}$, which is an on site interaction between the atoms of the 
two species. As customary, we introduce the definition of the population imbalance 
of each species as, $\hat{z}_a = (\hat{n}_{a,L}-\hat{n}_{a,R}) / N_a$ and 
$\hat{z}_b = (\hat{n}_{b,L}-\hat{n}_{b,R})/ N_b$. 

\begin{figure*}                        
\includegraphics[width=1.4\columnwidth,angle=-90]{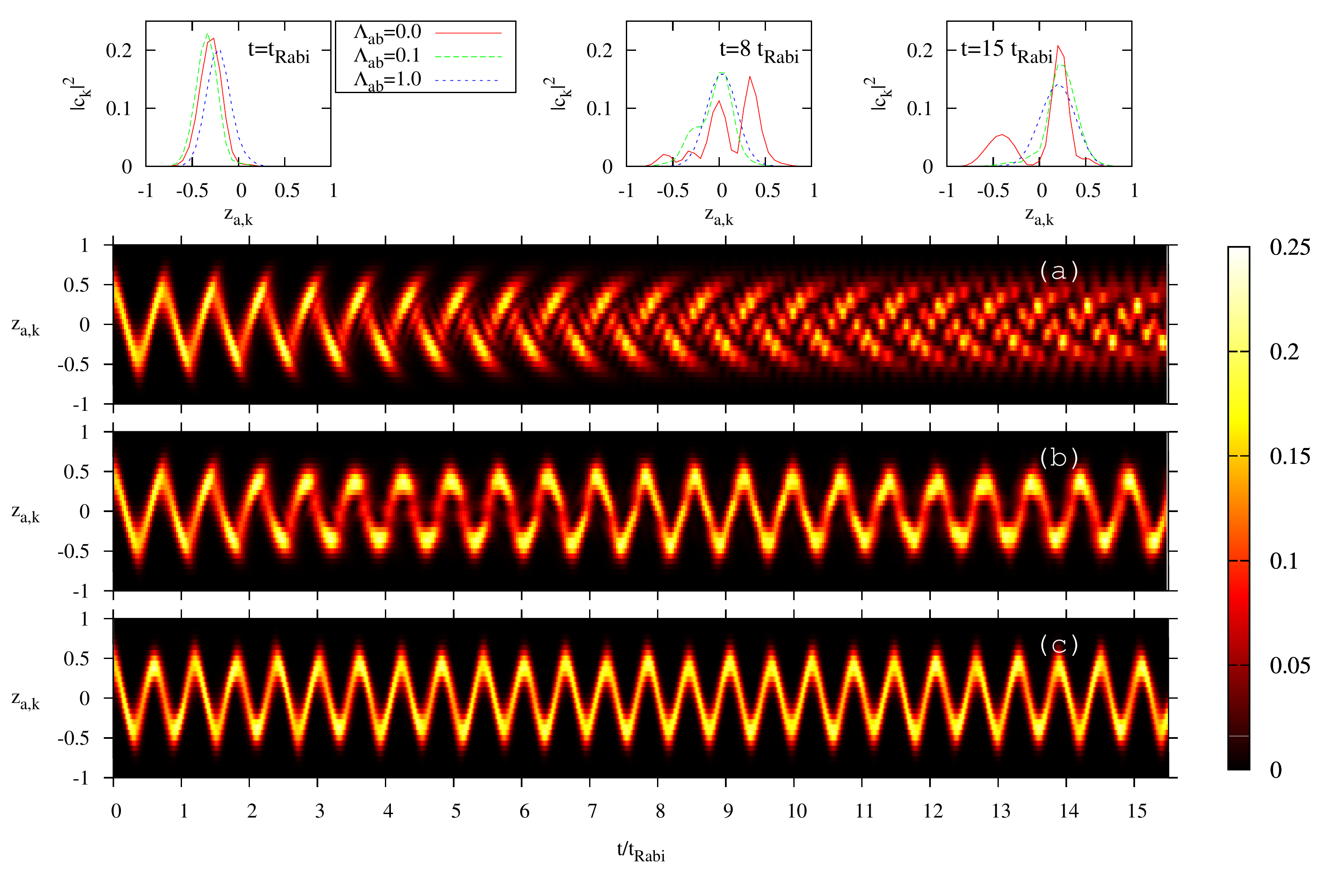}
\caption{(Color online)  
Evolution of the distribution of Fock coefficients, $|c_k|^2$ in Eq.~(\ref{eq:fock}) 
for the quantum gas $a$ as a function of time. We show three different values of 
the coupling $\Lambda_{ab}=0$ (a), $\Lambda_{ab}=0.1$ (b), and $\Lambda_{ab}=1.0$ (c). 
The insets depict the distribution $|c_k|^2$ (linked by a line) for three specific 
times $t=t_{\rm Rabi}$, $8\, t_{\rm Rabi}$, and $15\, t_{\rm Rabi}$ and the three 
different couplings. In all cases $N_a=N_b=30$, $J_a=J_b=1$ and $\Lambda_a=\Lambda_b=1$. 
The initial states are described in the text, $\langle z_a\rangle=z_b=0.4$ and 
$\phi_a=\phi_b=0$. The value of $|c_k|^2$ (color) is plotted as a function of the 
imbalance of each Fock state, $z_{a,k}=(2k/N_{a}-1)$, $k=0,\dots,N_a$. 
\label{fig1}}
\end{figure*}

The quantum ~\cite{vardi14} and classical (fully condensed) 
dynamics~\cite{Ashhab02,diaz09,satija09,mazzarella10,naddeo10}, of this 
model have been previously studied comparing different dimensional 
reductions of the three-dimensional equations ~\cite{mele11}. Also, measure 
synchronization~\cite{MS99} has been studied in the classical~\cite{Tian14} 
and full quantum case~\cite{Qiu14}. 

\subsection{Hybrid, quantum and classical, description of the system}

Our main interest is to study the combined evolution of $a$ and $b$. In particular, 
one of our aims is to discern whether the coupling to a condensed system, $b$, 
will enhance the degree of condensation of the ultracold gas, $a$. To answer this 
question, $b$ is assumed to be condensed at all times. 
Physically, this situation could be attained if the $b$ 
component has a large enough number of atoms. We thus neglect the quantum 
fluctuations of the $b$ cloud. In practice this is done by replacing the 
operators by $c-$numbers, $b_L = \sqrt{n_{b,L}} e^{\phi_{b,L}}$, 
$b_R = \sqrt{n_{b,R}} e^{\phi_{b,R}}$. We define the phase difference as 
$\phi_{b}=\phi_{b,R} -\phi_{b,L}$. 

The Hamiltonian is thus written as, $ \hat{H}=\hat{\cal H}_a+{H}_b+\hat{H}_{ab}$, 
with 
\beqa
H_b&=& -2 J_b\sqrt{n_{b,L} n_{b,R}} \cos \phi_b   \nonumber\\&+&
{U_b\over2} \left( n_{b,L} (n_{b,L}-1)+n_{b,R} (n_{b,R}-1) \right), \nonumber\\
\hat{H}_{ab} &=& {U_{ab}\over2}(\hat{n}_{a,L}- \hat{n}_{a,R})(N_bz_b) \,.
\label{eq:qc}
\eeqa
where $H_b$ is the $c-$number version of $\hat{\cal H}_b$.
To study the time evolution we solve the following coupled set of equations, 
\beqa
i \partial_t |\Psi_a \rangle &=& \hat{H}(t)  |\Psi_a\rangle, \label{eq:full}\\
\dot{z}_{b} &=& - 2J_{b}\sqrt{1-z_{b}^2} \sin\phi_b, \nonumber \\
\dot{\phi}_{b}&=& 2J_{b}{\Lambda_{b}} z_{b} + 2J_{b}{ z_{b}\over \sqrt{1-z_{b}^2}} \cos \phi_{b} 
+2J_{b}\Lambda_{ab} \langle \hat{z}_{a} \rangle(t)   \,.\nonumber
\eeqa
where we have introduced the dimensionless ratios, $\Lambda_a\equiv N_aU_a /(2J_a)$, 
$\Lambda_{b}\equiv N_{b}U_{b} /( 2J_{b})$, and $\Lambda_{ab} \equiv N_{a}U_{ab} /( 2J_{b}) $. 

The time dependence of $\hat{H}$ stems from the time dependence of $z_b$ and $\phi_b$. 
This set of equations is obtained considering the $b$ subsystem as a classical 
parametric driving for the $a$ subsystem and including the feedback effect 
of $a$ on $b$. Neglecting quantum fluctuations in one field (then classical), 
still retaining feedback effects due to the interaction with the other (fully quantum), 
is reminiscent of the `time-dependent parametric approximation'~\cite{TDPA1}, 
used to describe large quantum fluctuations in a convectively unstable signal 
of an optical parametric oscillator, when the pump field is approximated by 
a classical field while the signal is quantum~\cite{TDPA2}. Furthermore, here we 
are neglecting the effect of the quantum fluctuations of $\hat{z}_{a}$ in 
evaluating the dynamical evolution of $\phi_b$, as we approximate $\hat{z}_{a}$ by 
its expectation value.  

The coupled system of Eqs.~(\ref{eq:full}) is solved in the following way. 
We use a fourth-order Runge-Kutta routine to integrate the differential equation for $z_b$ and 
$\phi_b$ coupled to a unitary truncation of the Schr\"odinger equation 
for $|\Psi_a\rangle$ of the form, 
\beq
|\Psi_a(t+\Delta t)\rangle = 
\left( 1+ i {\Delta t\over 2\hbar}\hat{H}(t) \right)^{-1} \; 
\left(1- i{\Delta t\over2\hbar}\hat{H}(t)\right) \, |\Psi_a(t)\rangle\,,
\label{eqfor}
\eeq
with $\Delta t \approx 0.0002\,t_{\rm Rabi}$.
\begin{figure}[t]
\includegraphics[width=1.3\columnwidth,angle=-90]{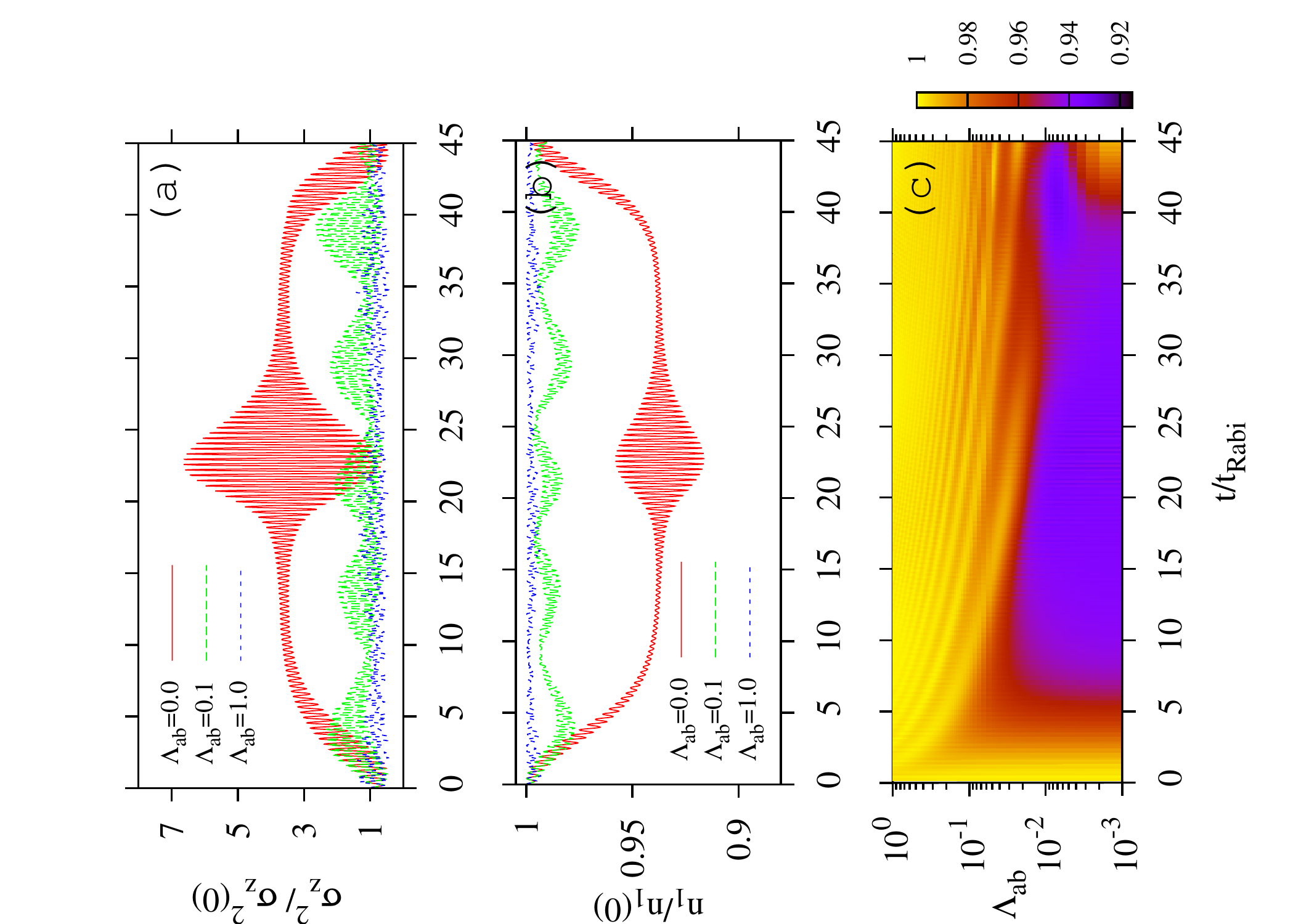}
\caption{(Color online) 
Evolution of two relevant many-body properties for different values of 
the interspecies coupling, $\Lambda_{ab}$. Panels (a) and (b) depict the dispersion of the 
population imbalance of the $a$ cloud, $\sigma_z^2$, and the condensed 
fraction of the $a$ system, $n_1$, respectively, for the values considered 
in Fig.~\ref{fig1}, $\Lambda_{ab}=0$, $\Lambda_{ab}=0.1$, and $\Lambda_{ab}=1.0$.
Panel (c) scrutinizes the evolution of the condensed fraction for a broader  
range of values of $0<\Lambda_{ab}<1$. All other parameters are the same as in 
Fig.~\ref{fig1}. Note that the initial condensed fraction, $n_1(0)=1$, so that 
the plotted values are $n_1$. 
\label{fig2}}
\end{figure}

The solutions of Eq.~(\ref{eq:full}) are numerically found to conserve the average 
energy, $\langle \Psi_a |\hat{H}(t) |\Psi_a\rangle\simeq \langle \Psi_a |\hat{H}(0) |\Psi_a\rangle $ 
in all the calculations reported in this paper. It is worth emphasizing that the 
coupling between $a$ and $b$ cannot simply be regarded as a driving term for $a$. 
Conservation of the total average energy implies on average a transfer of energy 
between the $a$ and $b$ subsystems. The dynamics is thus radically different from 
the case of a driven single component Josephson junction~\cite{Driven13,Driven10}, 
which for instance would occur if $z_b(t)$ in our description was replaced by a 
periodic function. In such case, energy would not be conserved and subsystem $a$ 
would gradually increase its energy. In our formulation the exchange of energy 
allows for the $mutual$ synchronization of both species 
and differs from entrainment as will be described in the following section.

Using the Fock basis of the 
$N_{a}+1$ dimensional space, $|n_{a,L},n_{a,R}\rangle=\{|N_a,0\rangle, \dots, |0,N_a\rangle\}$, 
the most general $a$ state is written as,
\begin{equation}
|\Psi_a\rangle= \sum_{k=0}^{N_{a}} c_k \ |k,N_{a}-k \rangle\,.
\label{eq:fock}
\end{equation}
All many-body properties of the state are computed from the $c_k$s, e.g. 
the average population imbalance of the $a$ cloud reads, 
\beq
\langle \hat{z}_a\rangle = \sum_{k=0}^{N_a}  |c_k|^2 z_{a,k}
\eeq
with $z_{a,k}=(2k/N_{a}-1)$. 
The degree of condensation of the cloud is 
given by its condensed fraction, i.e. the largest eigenvalue, $n_1$, 
of the single-particle density matrix $\rho^a(t)$, 
$\rho_{ij}^a(t)= {1\over N_a}\langle \Psi_a(t) |\hat{a}^\dagger_i \hat{a}_j| \Psi_a(t) \rangle$, $i,j=L,R$.  
The condensed fraction is also referred to as single-particle coherence.
With this normalization, a fully condensed cloud of $a$ would 
correspond to $n_1=1$ and $n_2=0$. In the following sections we 
discuss the time evolution of $n_1$ for different initial conditions 
and couplings.

\section{Coupled quantum and condensed dynamics}
\label{sec:dyn}

In this section we will describe how the appearance of synchronization 
in the combined evolution is found to be directly related to a 
coherent, in the sense of not dephased, evolution for the $a$ subsystem. 

\subsection{From dephasing to coherent evolution}

Let us first exemplify our discussion with one specific configuration. 
We will choose as initial state a condensed quantum state for $a$, 
i.e. all $a$ atoms populate the same single particle state 
$1/\sqrt{2}(\cos(\theta_a/2) a^\dagger_L + e^{i\phi_a} \sin(\theta_a/2)a^\dagger_R)|{\rm vac}\rangle$. 
The many-body state reads,  
\begin{eqnarray}
 \left|\Psi_a(\theta_a,\phi_a)\right>&=& \sum_{k=0}^{N_a}
\genfrac{(}{)}{0pt}{}{N_a}{k}^{1/2}[\cos(\theta_a/2)]^{k}
                 [\sin(\theta_a/2)]^{N_a-k}
                 \nonumber\\
                 &\times& e^{i(N_a-k)\phi_a}| k, N_a-k \rangle\;.
\end{eqnarray}
In particular we will take $\cos(\theta_a)=\langle \hat{z}_a\rangle(t=0)=0.4$ 
and $\phi_{a}=0$. To emphasize the effect of the coupling term we choose similar 
conditions for $b$, $z_{b}=0.4$, $\phi_{b}=0$, $N_{a}=N_{b}=30$, 
$J\equiv J_a=J_b=1$ and $\Lambda_{a}=\Lambda_{b}=1$. The Rabi time is $t_{\rm Rabi}=\pi/J$, and the 
Rabi frequency is $\omega_{\rm Rabi}=2 \pi /t_{\rm Rabi} = 2J$ 

\begin{figure}[t]
\includegraphics[width=8cm,angle=0]{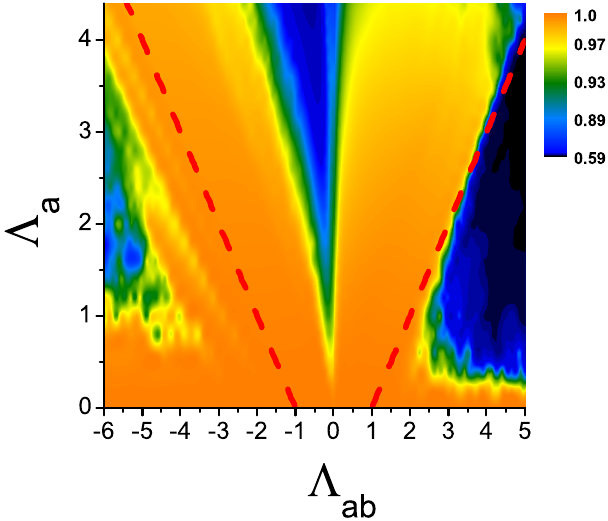}
\caption{(Color online) Time averaged condensed fraction 
$\overline{ n_{1}}= \int_0^T n_1(t)dt$ with $T=15\,t_{\rm Rabi}$ as a 
function of $\Lambda_{ab}$ and $\Lambda_{a}=\Lambda_b$. The initial 
condition is again fully condensed for the $a$ component with 
$\langle z_{a}\rangle (0)=z_{b}(0)=0.4$. All other parameters are as 
in Fig.~\ref{fig1}. The dashed red lines correspond to 
$1+\Lambda_a-\Lambda_{ab}=0$ and $1+\Lambda_a+\Lambda_{ab}=0$, 
see text for details. Notice that the scale in the color bar is 
nonlinear, and it spans all the values of the data.
\label{fig3}}
\end{figure}

In absence of coupling between $a$ and $b$, $\Lambda_{ab}=0$, the quantum 
system $a$ with a non-zero initial population imbalance evolves with time 
in a well studied fashion~\cite{Milburn97,jame05,bruno10}. Due to the 
atom-atom interactions which dephase the different Fock components, the 
initial distribution of $c_k$ evolves in time deforming its initial shape
(see Fig.~\ref{fig1}(a)). For the first oscillations, up to $t\simeq 5\,t_{\rm Rabi}$ 
the wave packet remains mostly unchanged, which in turn is also reflected 
in the fact that the $a$ component remains essentially condensed 
$n_1\gtrsim 0.98$ (see Fig.~\ref{fig2}(b)). For larger times, 
$t\gtrsim 10\,t_{\rm Rabi}$, the original shape is lost, the 
$a$ component is no-longer in a coherent quantum state, and 
thus the condensed fraction drops below $0.95$. Interaction 
among atoms $a$ is thus seen to decrease the degree of condensation 
of the subsystem fairly early. Directly related is the increase in the 
uncertainty on the particle number difference, 
$\sigma_z^2 = \langle \hat{z}_a^2\rangle - \langle \hat{z}_a\rangle^2$, shown 
in Fig.~\ref{fig2}(a), also appreciable in Fig.~\ref{fig1}(a).

\begin{figure}[t]
\includegraphics[width=0.95\columnwidth,angle=0]{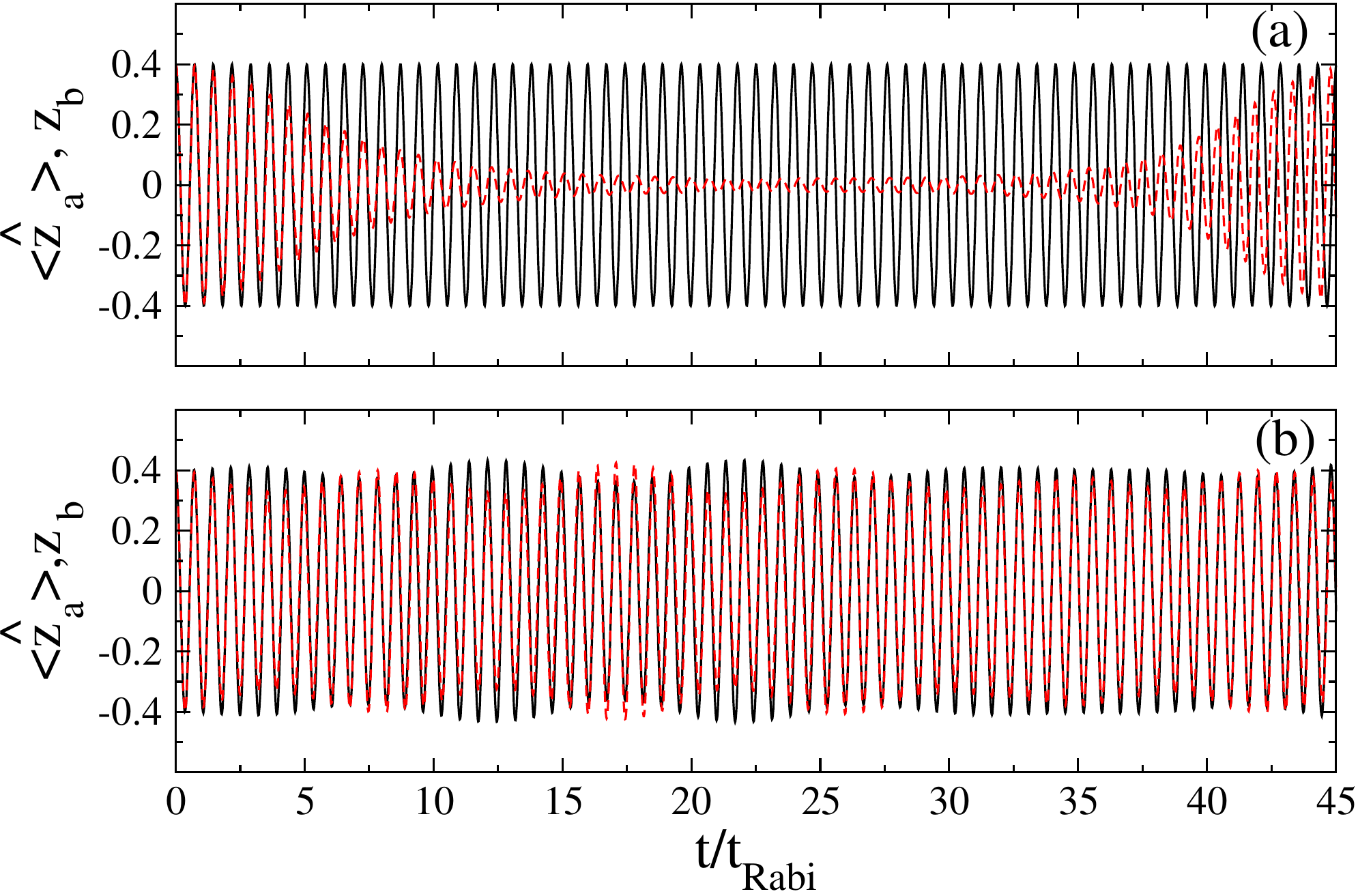}
\caption{
(Color online) 
Evolution of $\langle \hat{z}_{a} \rangle (t)$ (dashed-red), and $z_{b}(t)$ (solid-black) 
for two different values of the coupling (a) $\Lambda_{ab}=0$ and (b) $\Lambda_{ab}=0.1$. 
All other parameters are as in Fig.~\ref{fig1}.
\label{fig4}}
\end{figure}

When coupling the quantum system $a$ to the condensed one, $b$, a distinctive dynamics
is found. The dephasing due to the atom-atom interaction disappears 
and the quantum system remains condensed for longer times. As seen in 
Fig.~\ref{fig1}(b) for $\Lambda_{ab}=0.1$ the distribution of $|c_k|^2$ 
remains closer to a displaced binomial one, which again reflects in a much 
larger condensed fraction (see Fig.~\ref{fig2} and smaller insets in 
Fig.~\ref{fig1}). Already 
with this fairly small value of $\Lambda_{ab}$ we find a substantial increase 
in the condensed fraction, which is now at all times larger than 0.97. Further 
increasing $\Lambda_{ab}$, the effects are enhanced: The quantum system remains 
close to condensed for long times, see Fig.~\ref{fig1}(c), and the distribution 
of $|c_k|^2$ thus evolves, keeping its original shape. The latter is also 
reflected in $\sigma_z^2$, see Fig.~\ref{fig2}(a), which is found to remain almost 
constant in the coupled case. Effectively, the coupling 
to the condensed gas removes the dephasing effects due to the atom-atom 
interaction, obtaining an almost interaction-free evolution of the quantum 
system. 

This phenomenon is persistent in a broad range of parameters. 
For $\Lambda_{ab}\gtrsim 0.01$ and maintaining similar initial conditions, the 
quantum evolution is essentially coherent, as shown in Fig.~\ref{fig2} (b) and ~\ref{fig2}(c). 

Varying the values of $\Lambda_a=\Lambda_b$ and $\Lambda_{ab}$ one finds the 
following picture. One can easily prove that within our formalism for $\Lambda_a=0$, 
irrespective of the value of $\Lambda_{ab}$ the condensed fraction of $a$ remains 1. This 
behavior survives for small values of $\Lambda_{a}\lesssim 0.5$, for which a high 
degree of condensation for $a$ is also found (see Fig.~\ref{fig3}). 
For $\Lambda_{a}\gtrsim 0.5$, increasing $\Lambda_{ab}$ the condensed fraction 
of the system is found to remain essentially constant in time (see Fig.~\ref{fig3}). 
Thus, the coupling between $a$ and $b$ increases the coherence of the $a$ system. However,
for larger values of $\Lambda_{ab}$, in particular for $\Lambda_{ab}\ge \Lambda_{a}+1$ we 
observe a decrease of the condensation of $a$ (see Fig.~\ref{fig3}). This can be 
understood from the linear stability analysis of the classical equations around 
$z_a, z_b \ll 1$. In this case, one of the two natural modes~\cite{mele11}
$\omega_2=\omega_{\rm Rabi} \sqrt{1+\Lambda_a-\Lambda_{ab}}$ becomes unstable if 
$\Lambda_{ab} > \Lambda_a+1$, which induces decoherence of the $a$ system.

A similar picture is obtained in the attractive interspecies interaction 
case, $\Lambda_{ab}<0$ (see Fig.~\ref{fig3}). In particular, for values 
of $\Lambda_a<4$ we see that increasing the value of $|\Lambda_{ab}|$ 
the condensation of the $a$ cloud is increased. Further increasing $|\Lambda_{ab}|$, 
as observed for $\Lambda_a\simeq 1.5$, the degree of condensation decreases. 
The boundary of the classical stability region obtained for small values of the 
imbalance is found by imposing 
$\omega_1=\omega_{\rm Rabi} \sqrt{1+\Lambda_a+\Lambda_{ab}}$ to be real, 
$\Lambda_{ab} > -\Lambda_a-1$.

\subsection{Hybrid synchronization}
\label{sec:hs}

We have described how the coupling between the subsystems prevents 
the $a$ subsystem from fragmenting during the time evolution for 
certain coupling values. Now we show how this effect is directly 
connected to the appearance of synchronization between properties 
of both subsystems. This synchronization, to which we refer as 
hybrid, stemming from the hybrid nature of our coupled system, 
manifests itself in average properties. 

\begin{figure}[t]
\includegraphics[width=0.9\columnwidth,angle=0]{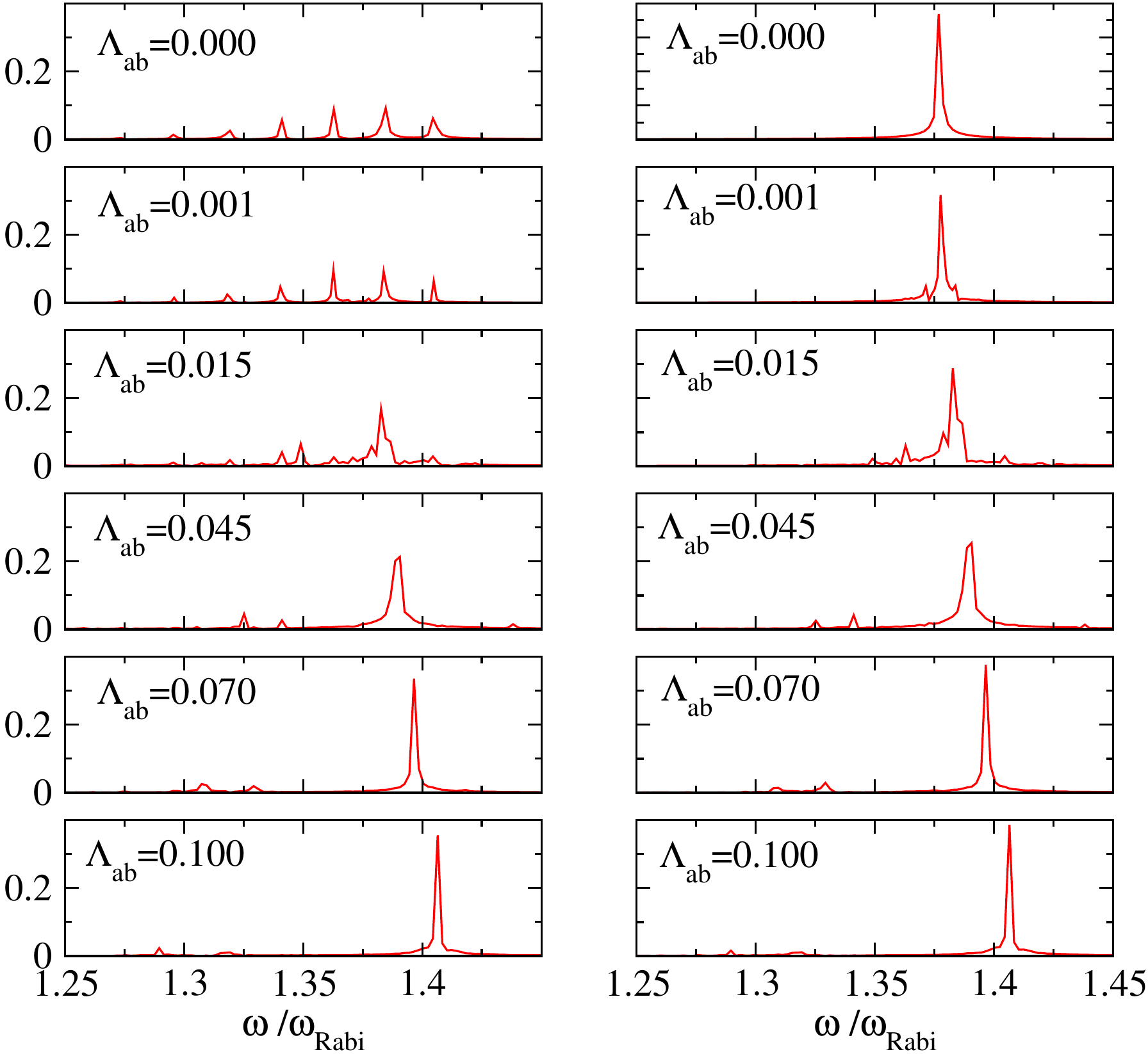}
\caption{(Color online) Absolute value of the frequency spectrum  $z(\omega)$ 
(Fourier Transform of the average population imbalances), for 
different values of the coupling $\Lambda_{ab}$. The frequency spectrum of the quantum result, $\langle  \hat{z}_{a} \rangle(\omega)$ is given in the left panels. $z_b(\omega)$ is depicted in the right panels. The frequency spectra are obtained from time series up to $T_{\rm Max}=585\; t_{\rm Rabi}$. }
\label{fig5}
\end{figure}

In the non-coupled case, Fig.~\ref{fig4} (a), the population imbalance 
of the condensed subsystem, $z_b$, is fully periodic~\cite{Smerzi97}. The 
frequency seen in the figure, $\omega \simeq 1.38 \, \omega_{\rm Rabi}$ is close to the 
one obtained linearizing around the $z_b=0$ fixed point, 
$\omega=\omega_{\rm Rabi} \,\sqrt{1+\Lambda_b} \simeq \omega_{\rm Rabi} \,\sqrt{2}$. The population imbalance of subsystem $a$, 
$\langle \hat{z}_a\rangle$, features characteristic collapses 
and revivals, which are also present in the condensed fraction and population imbalance dispersion shown in Fig.~\ref{fig2}~\cite{Milburn97}. 

\begin{figure}[t]
\includegraphics[width=0.95\columnwidth,angle=-90]{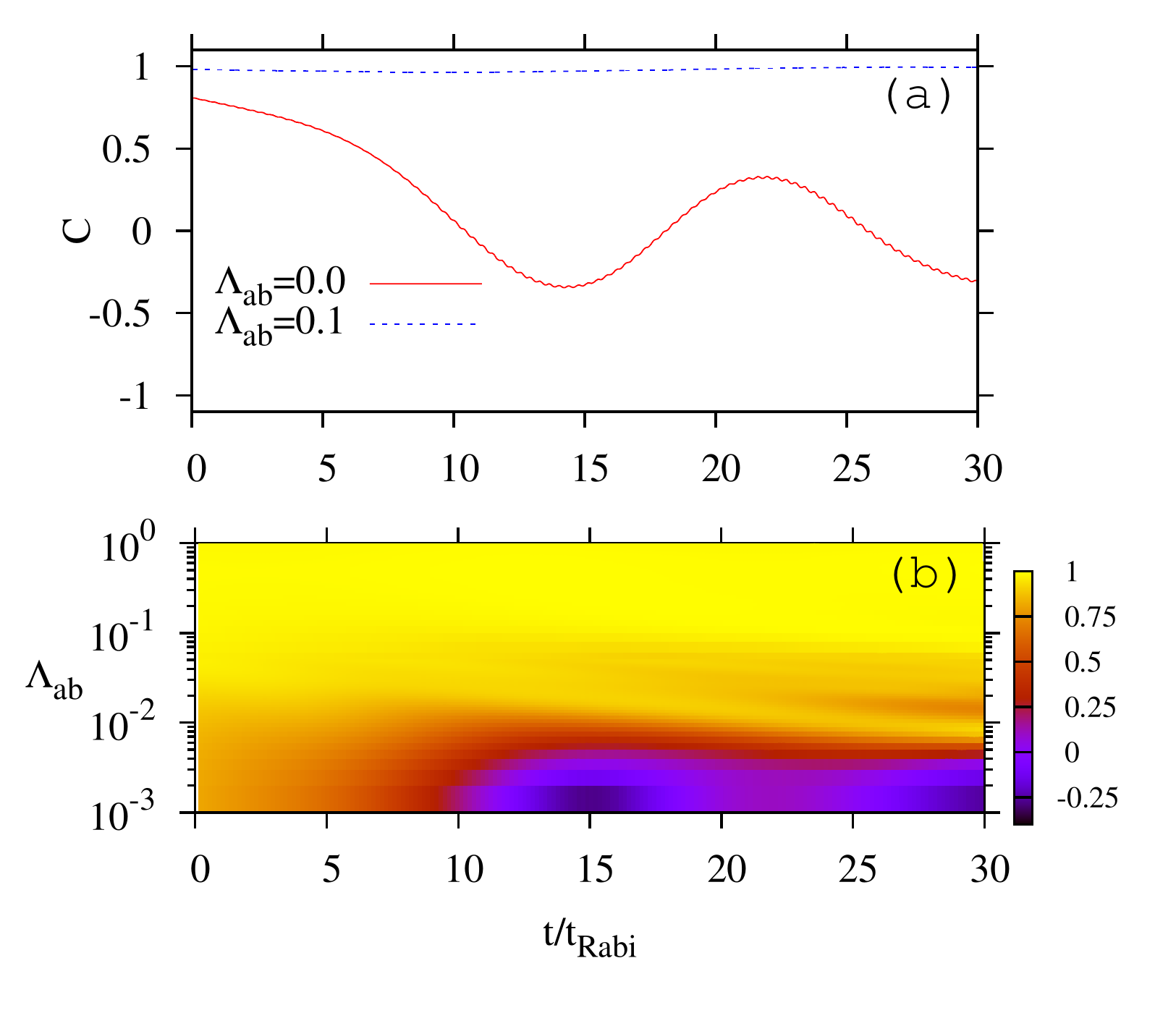}
\caption{(Color online) Time correlation coefficient $C_{\langle z_a\rangle,z_b}$ 
defined in Eq.~(\ref{eq:c}) versus the coupling strength $\Lambda_{ab}$. 
Panel (a) corresponds to two specific values of $\Lambda_{ab}=0$, 
$\Lambda_{ab}=0.1$, while (b) contains the time evolution of $C_{<z_{a}>,z_{b}}$ 
for a broader set of $\Lambda_{ab}$. $\Delta t=14 \,t_{Rabi}$. All other 
parameters are as in Fig.~\ref{fig1}.
\label{fig6}}
\end{figure}

Coupling both subsystems, Fig.~\ref{fig4} (b), both signals are found to 
be much more correlated. To quantify this we compare in Fig.~\ref{fig5} 
the frequency spectra of these signals for different 
values of  $\Lambda_{ab}$. In the uncoupled case the quantum signal is 
found to have several peaks around the same frequency $\omega \simeq 1.38 \,\omega_{\rm Rabi} $. 
The different equispaced peaks reflect the long evolvent seen in 
Fig.~\ref{fig4} (upper panel). They arise from the atom-atom interaction 
which makes the spectrum of the many-body Hamiltonian depart from the 
equispaced/harmonic case in the uncoupled case producing quantum 
revivals~\cite{Milburn97,Don04}. 

As $\Lambda_{ab}$ is increased, the spread of the peaks in the quantum case 
is reduced. For $\Lambda_{ab}=0.1$ the Fourier decompositions of both 
signals are very similar, showing a large peak at $\omega t_{\rm Rabi} \simeq 1.405$. For this value of 
$\Lambda_{ab}$, $a$ is mostly condensed and the classical description of the 
full system should approximately hold~\cite{Ashhab02,diaz09}. Indeed, the found 
frequency is reproduced by the classical equations~\cite{diaz09}. For this particular 
case of similar initial conditions of $a$ and $b$, the classical description of 
the binary mixture predicts, linearizing around $z \ll 1$, just one frequency, 
$\omega_1=\omega_{\rm Rabi} \sqrt{1+\Lambda_a+\Lambda_{ab}}=\sqrt{2.1}\,\omega_{\rm Rabi}$~\cite{diaz09}. 
The deviation observed is due to the departure from $z\ll 1$ of our 
initial conditions $\langle z_a \rangle (0) =z_b(0)=0.4$. 

The synchronization phenomenon, as seen in the Fourier analysis of Fig.~\ref{fig5} 
goes from several different frequencies for both subsystems in the uncoupled case, 
to a single major frequency in the coupled case. Thus the two subsystems 
get frequency-locked as the interaction is increased. It is also worth noting again that 
in our description subsystem $a$ can also remain condensed, but $b$ is not allowed 
to fragment. The resulting scenario is that the coupling induces condensation in 
subsystem $a$.

\begin{figure}
\includegraphics[width=0.95\columnwidth,angle=-90]{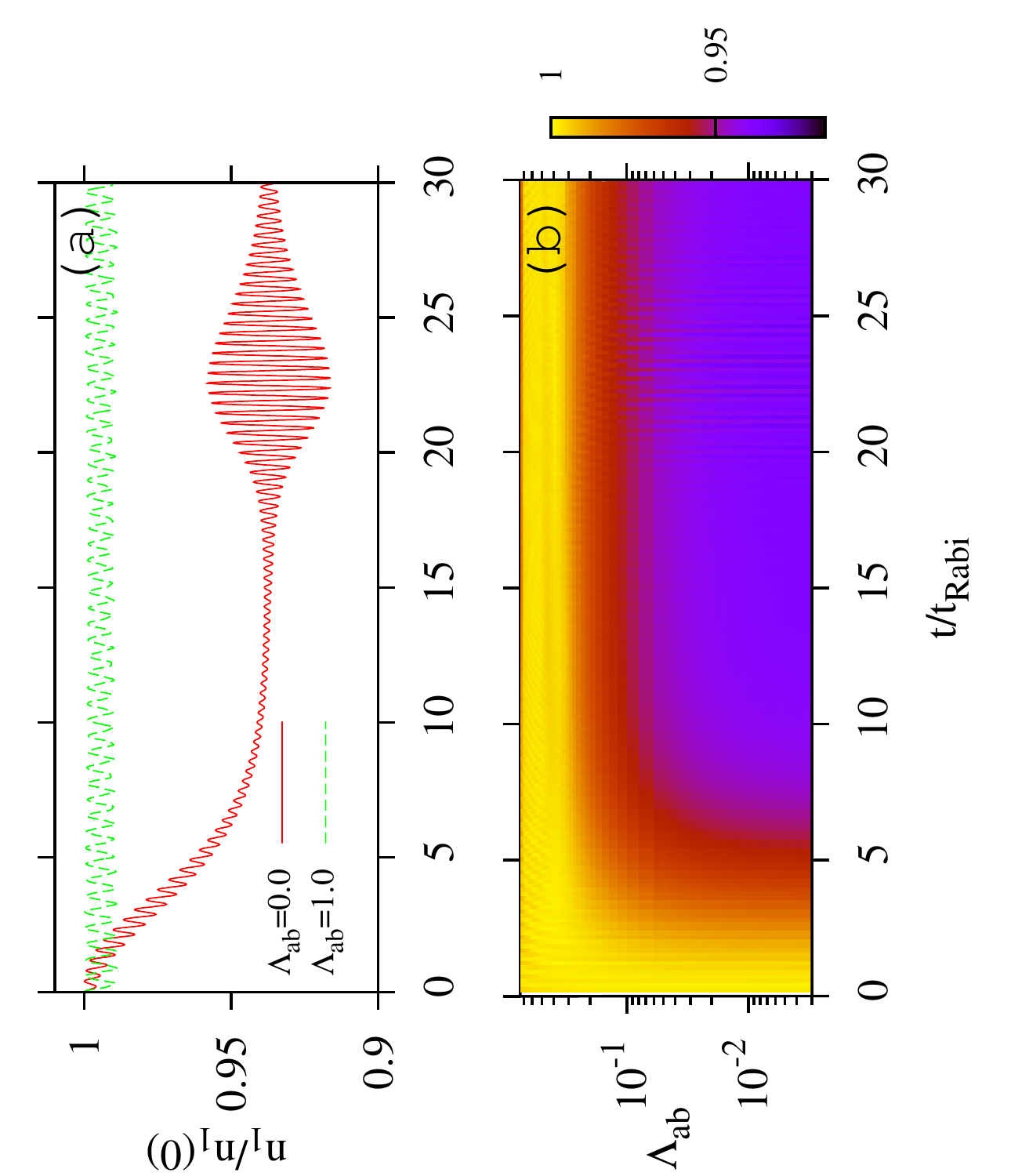}
\caption{(Color online) Evolution of single particle coherence $n_{1}$ for 
different values of $\Lambda_{ab}$. With unequal population $\frac{N_{b}}{N_{a}}=10$. 
In all cases we fixed $\Lambda_a=1$, $\Lambda_b=10\Lambda_a$, and $N_{b}=300$, 
$N_{a}=30$. The initial state and other parameters are as Fig.~\ref{fig1}. (a) 
corresponds to the two values of $\Lambda_{ab}$. (b) scrutinizes the same function 
for a more detailed range of values of $0<\Lambda_{ab}<1$.
\label{fig7}}
\end{figure}

\begin{figure}[t]
\includegraphics[width=0.95\columnwidth,angle=0]{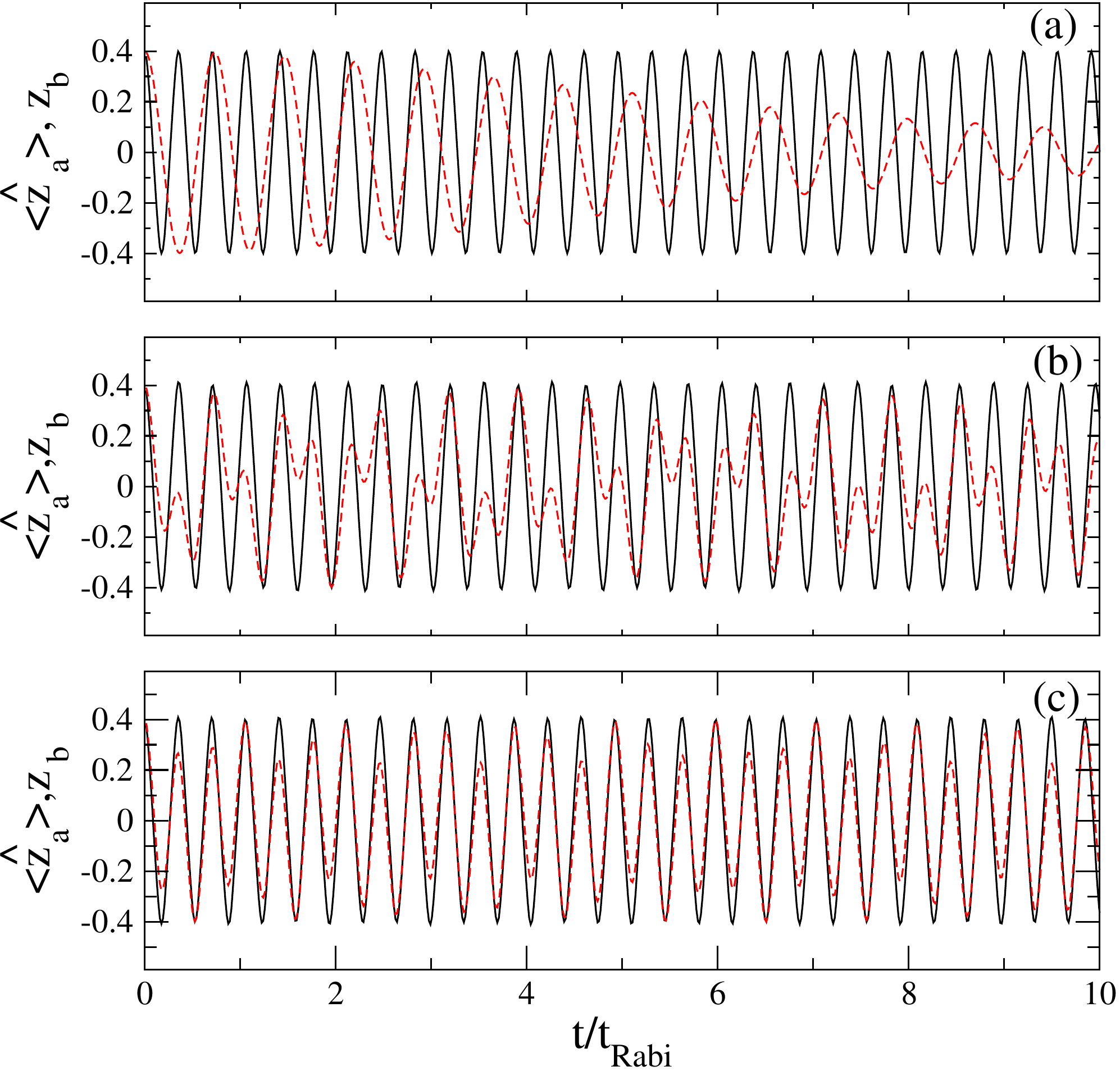}
\caption{
(Color online) 
Evolution of $\langle \hat{z}_{a} \rangle(t)$ (dashed-red), and $z_{b}(t)$ (solid-black) with  $\Lambda_b=10\Lambda_a$, $\Lambda_a=1$, and unequal populations $N_{a}=30$, $N_{b}=300$. Three different values of the coupling are chosen, with (a) $\Lambda_{ab}=0$, (b) $\Lambda_{ab}=0.1$, and (c) $\Lambda_{ab}=0.6$. 
All other parameters are as in Fig.~\ref{fig7}.
\label{fig8}}
\end{figure}

\begin{figure}
  \includegraphics[width=0.95\columnwidth,angle=-90]{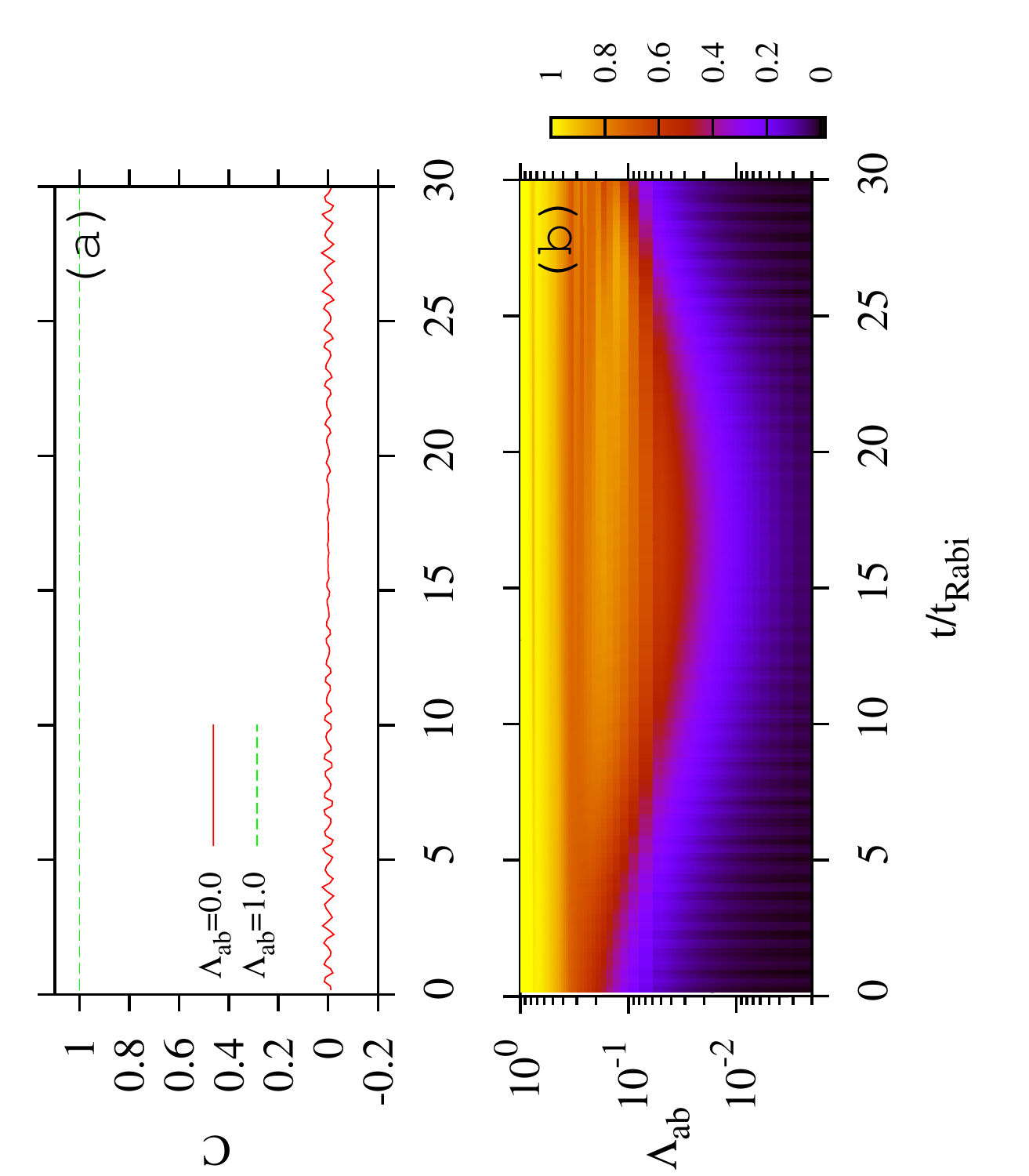}
\caption{(Color online) Evolution of $C_{\langle z_a\rangle,z_b}$ 
for different values of $\Lambda_{ab}$. With unequal population $\frac{N_{b}}{N_{a}}=10$. 
In all cases we fixed $\Lambda_a=1$, $\Lambda_b=10\Lambda_a$, and $N_{b}=300$, $N_{a}=30$.
The initial state and other parameters are as Fig.~\ref{fig1}. (a) corresponds to 
the two values of $\Lambda_{ab}$. (b) scrutinizes the same function for a more detailed 
range of values of $0<\Lambda_{ab}<1$.
\label{fig9}}
\end{figure}

\begin{figure}[t]
\vspace{10pt}
\includegraphics[width=0.9\columnwidth,angle=-90]{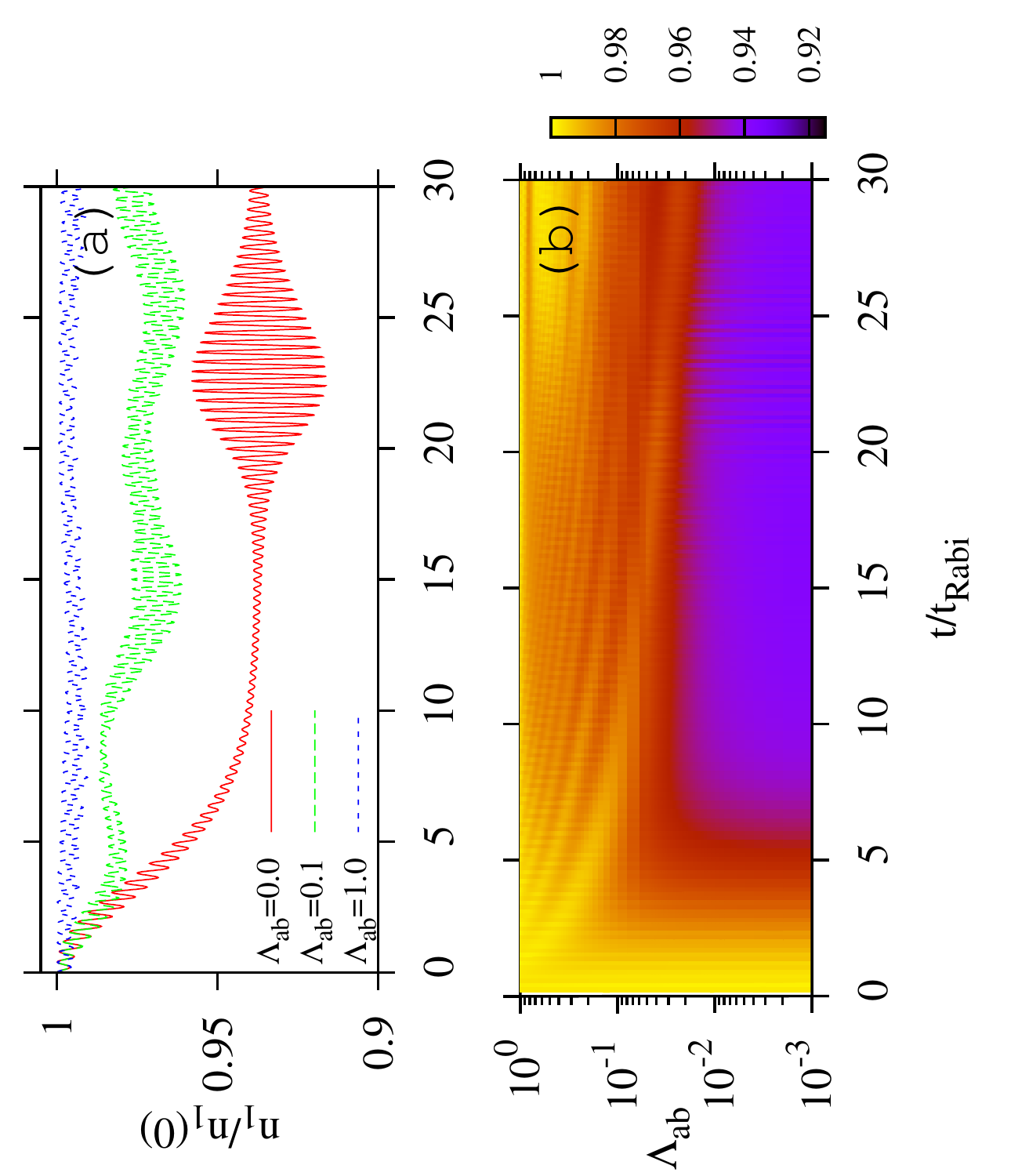}
\caption{(Color online) Evolution of the condensed fraction of 
subsystem $a$, $n_1$, as a function of time for different values
of the coupling $\Lambda_{ab}$. The upper panel corresponds to the three
values, $\Lambda_{ab}=0$, $0.1$, and $1.0$. The 
lower panel scrutinizes the same function for a more detailed range 
of values of $0<\Lambda_{ab}<1$. Here we consider different initial 
conditions, $\langle  \hat{z}_{a} \rangle(0)=0.4$, and $z_{b}(0)=0.2$.
\label{fig10}}
\end{figure}

\begin{figure}[t]
\vspace{10pt}
\includegraphics[width=8cm,angle=0]{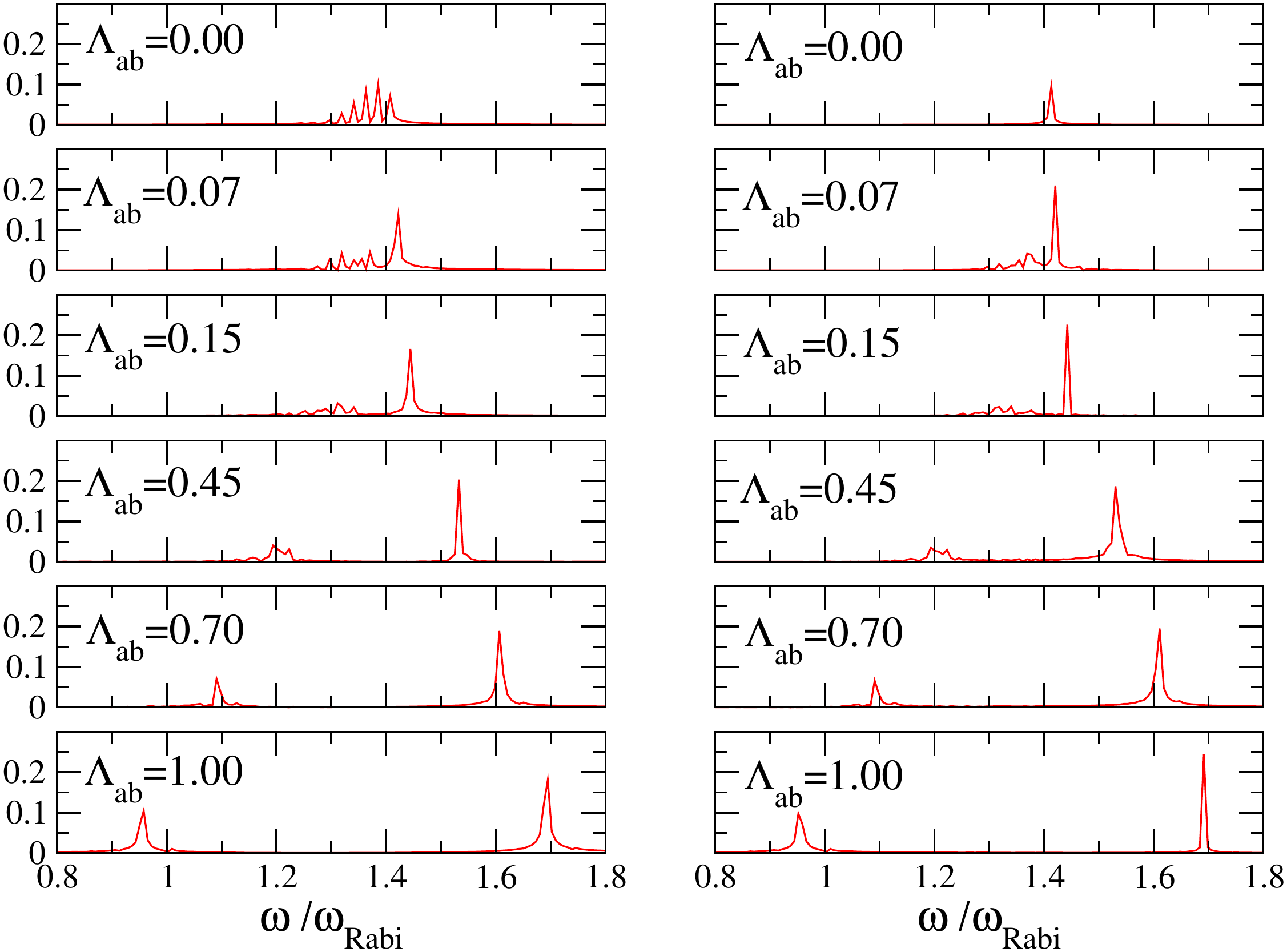}
\caption{(Color online) Absolute value of the frequency spectrum  $z(\omega)$  
(Fourier transform of the average population imbalances), for different values of the
coupling $\Lambda_{ab}$.  Here we consider different initial
imbalance in the two subsystems, $\langle  \hat{z}_{a} \rangle (0)=0.4$, and $z_{b}(0)=0.1$. The frequency spectrum of the quantum
result, $\langle  \hat{z}_{a} \rangle(\omega)$ is given in the left panels. $z_b(\omega)$
is depicted in right panels. 
The frequency spectra are obtained from time series up to  $T_{\rm Max}=140\, t_{\rm Rabi}$.}
\label{fig11}
\end{figure}

\begin{figure}[t]
\includegraphics[width=0.95\columnwidth,angle=0]{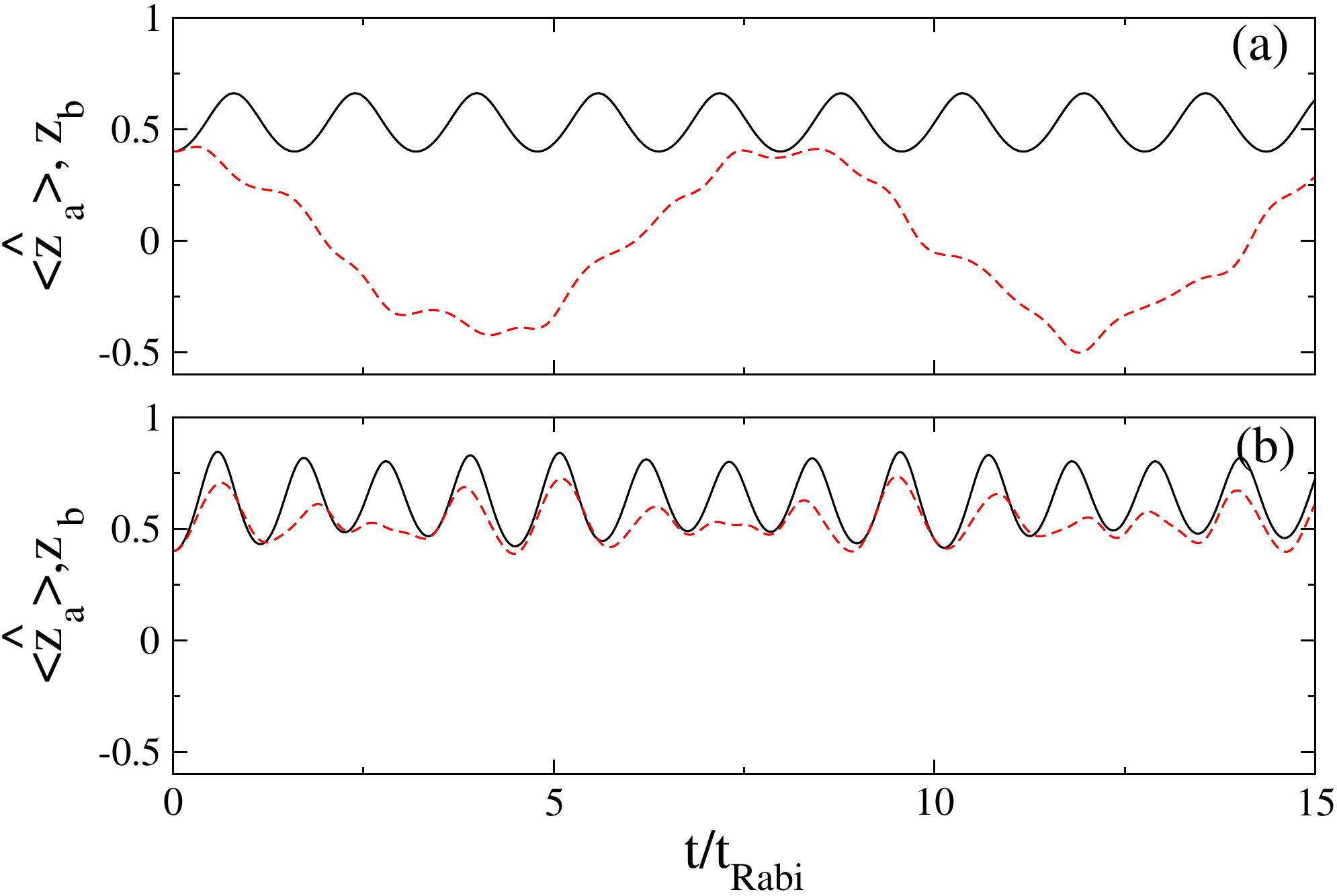}
\caption{
(Color online)
Evolution of $\langle \hat{z}_{a} \rangle (t)$ (dashed red), and $z_{b}(t)$ (solid black)
for two different values of the coupling (a) $\Lambda_{ab}=0$ and (b) 
$\Lambda_{ab}=0.2$.  The initial conditions are $\langle z_{a}\rangle (0)=z_{b}(0)=0.4$ 
and $\phi_a=\phi_b=\pi$ for the $\pi$-phase mode, with $\Lambda_{a}=\Lambda_b=1.2$, and 
$N_a=N_b=30$.
\label{fig12}}
\end{figure}

\begin{figure}
\includegraphics[width=0.95\columnwidth,angle=-90]{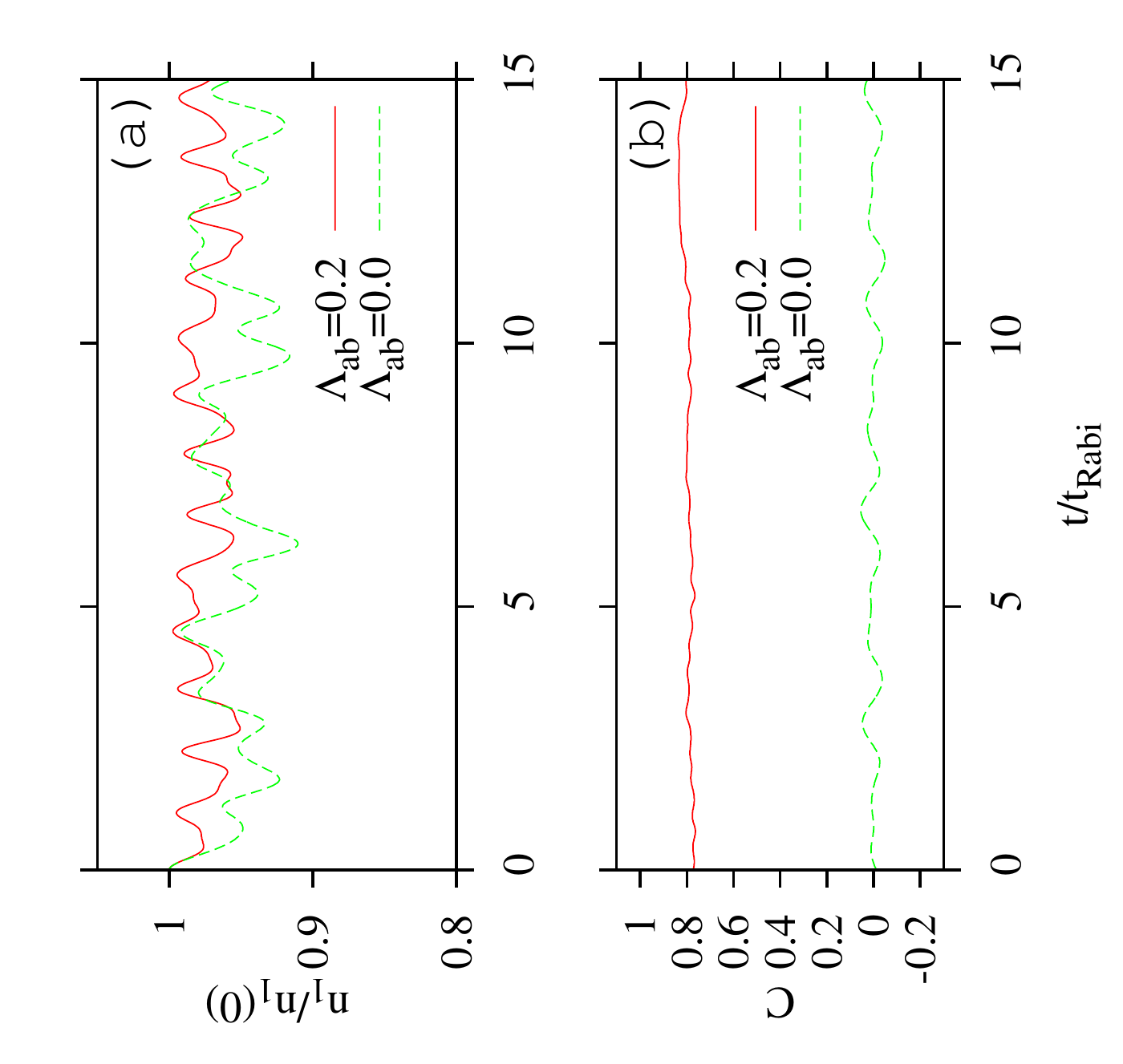}
\caption{
(Color online)
Evolution of  the condensed fraction of the $a$ system, $n_1$ (a), 
and time correlation coefficient $C_{\langle z_a\rangle,z_b}$ (b), for 
the values considered in Fig.~\ref{fig12}. The initial condensed 
fraction, $n_1(0)=1$.
\label{fig13}}
\end{figure}

In order to have a quantitative characterization of the synchronization, we calculate 
the time correlation coefficient $C$, which can be used to judge whether two time series 
are synchronized~\cite{Syn6,Zam2}. For two time signals, $f(t)$ and $g(t)$, it is defined as, 
\beq
C_{f,g}(t,\Delta t)={\overline{ \delta f(t) \delta g(t)} \over \sqrt{\overline{\delta f^2(t)} \ \overline{\delta g^2(t)} }}
\label{eq:c} 
\eeq
where the bar stands for a time average 
$\overline{f}(t)=\frac{1}{\Delta t}\int_{t}^{t+\Delta t}dt'f(t')$ with 
time window $\Delta t$ and $\delta f(t)=f(t)-\overline{f}(t)$. We 
choose a time window $\Delta t=14\,t_{\rm Rabi}$. 
For in phase (anti-phase) 
synchronization $C\sim 1$ ($-1$), while it equals zero for fully non-synchronized cases.

In Fig.~\ref{fig6}, we show $C$ as a function of time for different values of 
the coupling strength $\Lambda_{ab}$. The figure has a similar structure as that 
found when computing the condensed fraction of the same subsystem in Fig.~\ref{fig2}. 
The subsystem is seen to remain condensed for the same values of $\Lambda_{ab}$ for 
which the two signals are synchronized. 

For $\Lambda_{ab}<0$, we have shown that the results were 
similar to the $\Lambda_{ab}>0$ case, that is the condensation 
for the $a$ cloud is enhanced with enough coupling (see Fig.~\ref{fig3}). 
As occurred in the $\Lambda_{ab}>0$ case, the increase in condensation 
for $\Lambda_{ab}<0$ is also accompanied by an increase in the time 
correlation function between the 
average populations of both species, e.g. for $\Lambda_{ab}=-1.0$ with 
the same parameters and initial conditions as in Fig.~\ref{fig4}, 
$C\sim 1$ is reached.

\subsubsection{Effect of unequal populations}
 
The hybrid system we are considering, with $b$ remaining fully coherent during 
the time evolution, is justified if the number of $b$ atoms is 
large enough. Up to now we have discussed the case in which $N_a=N_b$, in order to 
isolate the effect of the inter-species coupling.

Now, we consider a system with unequal population $N_{b}/N_{a}=10$. With 
$N_{b}=300$, $N_{a}=30$, and $\Lambda_b=10\Lambda_a$, $\Lambda_a=1$, so that we have 
$U_{a}=U_{b}$. We keep the initial state and other parameters the same as 
in Fig.~\ref{fig1}. The results show a  picture very similar to that of 
the case with equal populations. In Fig.~\ref{fig7}, we show the evolution 
of single-particle coherence, $n_{1}$, for different values of $\Lambda_{ab}$.
We observe an increase of the condensed fraction as $\Lambda_{ab}$ is increased.  
This is similar to what has been shown in Fig.~\ref{fig2}. However, to make a 
more quantitative comparison of Fig.~\ref{fig2}(c) and  Fig.~\ref{fig7} (b), 
we will notice that for the case with unequal populations, in order to reach 
the same level of single particle coherence as for the case with $N_{a}=N_{b}$, 
a relatively larger $\Lambda_{ab}$ is needed now.

Fig.~\ref{fig8} shows the hybrid synchronization for this case ($N_{b}\neq N_{a}$). In the non-coupled case, Fig.~\ref{fig8} (a), the population imbalance of the condensed subsystem, $z_b$, is fully periodic. Compared with the equal population case, we notice that the frequencies of the two signals are very different. However, by coupling both subsystems with  $\Lambda_{ab}=0.1$, as seen in Fig.~\ref{fig8} (b), the signals are found to be more correlated. Further increasing $\Lambda_{ab}$, the two signals show synchronous dynamics due to this coupling effect[Fig.~\ref{fig8} (c)].

To quantify the hybrid synchronization, in Fig.~\ref{fig9} we show the evolution of $C_{\langle z_a\rangle,z_b}$, for different values of $\Lambda_{ab}$. With 
$\Lambda_{ab}=0$, Fig.~\ref{fig9}(a), the time correlation function is 
close to zero as expected. With  $\Lambda_{ab}=1$, Fig.~\ref{fig9}(a), 
the time correlation function is close to $1$ as hybrid synchronization 
occurs. By comparing Fig.~\ref{fig9}(b) with Fig.~\ref{fig6}(b), one sees 
that in order to reach hybrid synchronization in the case with unequal 
populations, a larger coupling strength $\Lambda_{ab}$ is needed.
This is a general feature of synchronization arising when the coupling between systems is large enough to overcome their detuning.
Indeed here the detuning between the two clouds increases with the difference between the populations and needs to be compensated by a larger reciprocal contact interaction.

\begin{figure}[t]
\includegraphics[width=0.99\columnwidth,angle=-90]{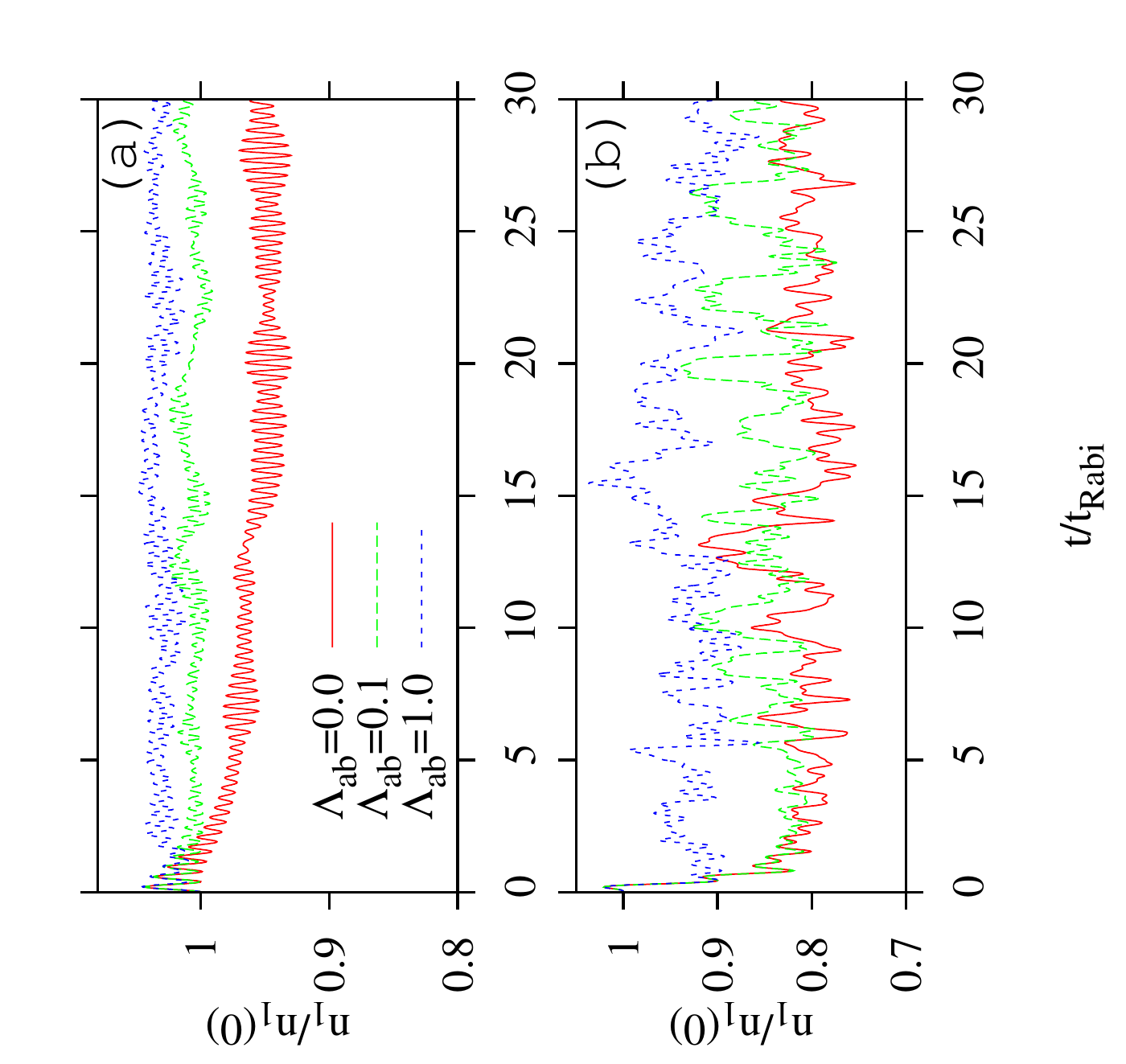}
\caption{(Color online) Evolution of the condensed fraction of the subsystem 
as a function of time for three different values of $\Lambda_{ab}$. In the 
upper panel the initial state is a squeezed state, with an initial condensed 
fraction $n_1(0)=0.9216$. In the lower panel, the initial state is the Fock 
state, $|21,9\rangle$, which has an average population imbalance of 0.4 and an initial 
condensed fraction of $21/30=0.7$. All other parameters are the same as in 
Fig.~\ref{fig1}. 
\label{fig14}}
\end{figure}

\begin{figure}[t]
\includegraphics[width=0.99\columnwidth,angle=-90]{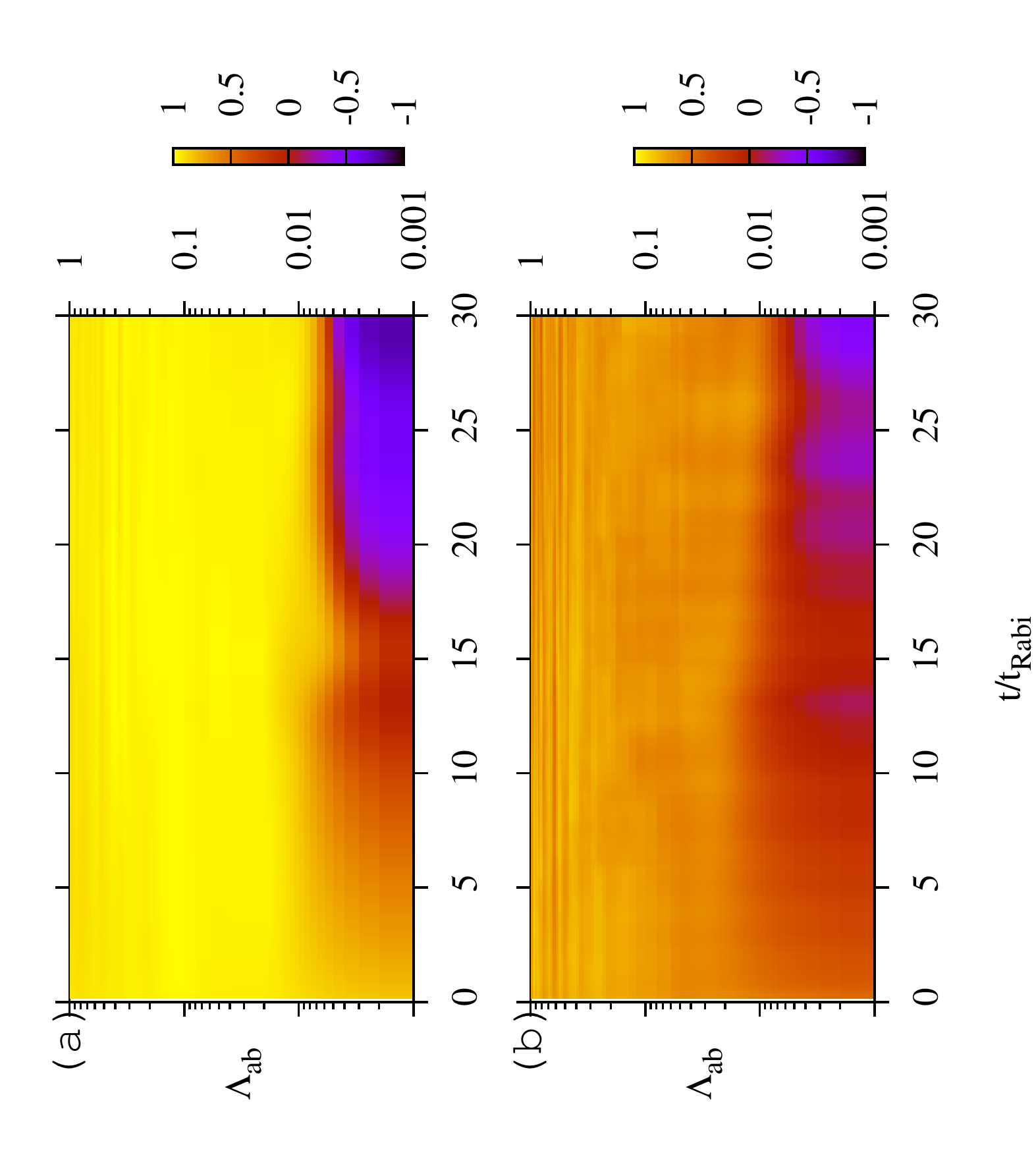}
\caption{(Color online) Time correlation coefficient $C_{\langle z_a\rangle,z_b}$ 
defined in Eq.~(\ref{eq:c}) versus the coupling strength $\Lambda_{ab}$ for the 
squeezed (upper panel) and Fock (lower panel) cases considered in Fig.~\ref{fig14}.  
\label{fig15}}
\end{figure}

\subsubsection{Effects of different initial preparation}

Up to now we have considered the same initial population imbalance 
for $a$ and $b$. Now we consider a more general case, in which the initial 
average population imbalance is not the same for both subsystems. 
In particular we will consider $\langle z_a\rangle=0.4$ and $z_b=0.2$, while 
all other conditions are as in Fig.~\ref{fig1}. The 
main finding described earlier is again found: Coupling $a$ and $b$ increases 
the coherence time of subsystem $a$. 
As described above, the fragmentation of subsystem $a$ which takes 
place during the uncoupled evolution, see Fig.~\ref{fig10}, decreases
as the coupling between $a$ and $b$ is increased. For values of 
$\Lambda_{ab}\simeq \Lambda_a=\Lambda_b$, $a$ remains almost fully condensed 
during the evolution. 

Let us analyze the Fourier decomposition of the evolution of the population 
imbalance. Following our previous discussion in the case of decoupled clouds, 
we find that $z_b$ has only one frequency, which now is closer to the 
one expected in the linear regime, $\omega\simeq \omega_{\rm Rabi}\sqrt{1+\Lambda_b}=
\sqrt{2}\,\omega_{\rm Rabi}$. The $a$ subsystem, as before has a number of peaks, whose spread
is related to the deviation from the equispaced spectrum (which in this case is 
smaller). As we couple the two subsystems, the spread in the $a$ subsystem 
disappears and two prominent frequencies appear both for $a$ and $b$. Already 
for $\Lambda_{ab}=0.45$, we have the same Fourier structure in both cases, signaling
the appearance of hybrid synchronization.

The net effect is that for a strong enough coupling both signals oscillate with 
the same frequency, which is different from the free frequencies. As found in the case of equal initial preparations, the fact that $a$ remains mostly condensed makes a fully 
classical description of the complete system plausible. Indeed, the 
two frequencies remaining in the coupled case are reproduced by the 
classical equations~\cite{diaz09}. In summary, for different preparation 
of initial imbalance, it is found that the frequency locking with a single
prominent peak appearing as shown in Fig.~\ref{fig5} is replaced
by frequency locking of more complex dynamics featuring several spectral components 
displayed in Fig.~\ref{fig11}. 
This result reminds us of what happens for two coupled classical systems, 
in which the coupling will induce measure synchronization (MS)~\cite{Tian14}. 
In classical MS, the coupled dynamics will exhibit quasi-periodic 
motions, such that the Fourier analysis of $z_{a}$ and $z_{b}$ shows 
many peaks rather than one. And in this very special case, the hybrid 
synchronization is accompanied by MS in the combined dynamics. We emphasize 
that in general these two phenomena do not need to arise together.

The overall picture as $\Lambda_{ab}$ is varied, see Fig.~\ref{fig10}, 
is similar to the case of equal initial average population imbalance 
(see Fig.~\ref{fig2}). The effect is slightly degraded, finding a lower
condensed fraction for similar values of $\Lambda_{ab}$ in the case 
of different initial average population imbalances. We have also 
considered different choices of the nonlinearity, i.e., 
$\Lambda_{a}=2$, $\Lambda_{b}=1$. This has a similar effect as the different 
preparation of initial imbalances. 

Up to now we have only considered hybrid synchronization around 
stable phase-space points with $\phi=0$. In this case, the classical 
description of the Josephson junction~\cite{Smerzi97} shows a single 
stable minimum for $z=0$ with repulsive interactions. A similar single 
solution, non-bifurcated, is found for initial preparations 
$\phi_a=\phi_b=\pi$ if ($\Lambda_a<1, \Lambda_b<1$). In this case, 
we find similar results as those reported above. A more involved situation
is found if we consider a bifurcated region of the phase space of each 
individual Josephson junction, for instance, $\phi=\pi$ and $\Lambda>1$. 
In this case, the classical description of the junction predicts a self-trapped 
regime~\cite{Raghavan}. To illustrate this dynamical regime, we have 
considered the initial condition $\langle z_{a}\rangle (0)=z_{b}(0)=0.4$ 
and $\phi_a=\phi_b=\pi$, with $\Lambda_{a}=\Lambda_b=1.2$, such that the 
classical description of each junction (uncoupled) would predict a self-trapped 
regime. In the non-coupled case, see Fig.~\ref{fig12} (a), the population 
imbalance of the condensed subsystem, $z_b$, is fully periodic and 
self-trapped. The population imbalance of subsystem $a$, 
$\langle \hat{z}_a\rangle$, features a much more complicated dynamics with no
self-trapping. Coupling both subsystems with $\Lambda_{ab}=0.2$, 
Fig.~\ref{fig12} (b), both signals are found to be much more correlated. 
The time correlation coefficient $C_{\langle z_a\rangle,z_b}$ goes from $\simeq 0$ 
for the uncoupled case, see Fig.~\ref{fig13}(b), to a value close to $0.8$ 
for the coupled case ($\Lambda_{ab}=0.2$). Simultaneously, the degree of 
condensation, similarly to the case of the $0$-phase mode discussed above, 
is found to increase with the coupling between the two systems, although the 
increase is less notable than in the zero phase case. 
Furthermore, we have tried different initial conditions for the 
$\pi$-phase mode in the bifurcated region of the classical phase space, and find out that due to the instability associated with the bifurcation~\cite{zib10}, the parameter space is much smaller compared with $0$-phase mode in achieving hybrid synchronization.

\subsubsection{Initially squeezed and Fock states}

In all previous results the initial state considered was a condensed many-body 
state, i.e. condensed fraction $n_1=1$. Thus, the effect we have described up 
to now is how by coupling the quantum to a condensed system the condensed 
fraction of the quantum state was found to get closer to 1. For that case, 
the coupling to the condensed state was thus helping the quantum system 
to remain coherent during the time evolution. 

In this section we broaden the set of initial states to consider squeezed 
and Fock states. Squeezed states are particularly useful as they can be used 
to improve the efficiency of interferometers made with ultracold atomic 
systems~\cite{kita93,gross10,riedel10}. In brief what we find is that 
coupling the quantum system to the condensed one has a similar effect 
as what was described before, i.e. the coupling mostly prevents the 
dephasing and thus makes the condensed fraction of the quantum system remain approximately constant with time. 

In Fig.~\ref{fig14} we consider similar conditions as in Fig.~\ref{fig1}, but 
with squeezed and Fock initial states. The squeezed initial state is built 
as, $c_k  = e^{ - \frac{{(k - k_0 )^2 }}{{2\sigma ^2 }}} /\sqrt {\sigma \sqrt \pi  },$ where 
$k_0$ sets the value of the population imbalance and $\sigma$ sets the squeezing 
of the state. The coherent state considered above has $\sigma=\sqrt{N_a/2}$. 
A smaller value of $\sigma$ provides a squeezed initial state. In the 
figure we have taken $\sigma=1$. The picture is very similar to the case of 
an initial coherent state. Increasing the coupling between the condensed and quantum 
subsystems the condensed fraction is seen to remain closer to its initial value (not $1$ 
in this case).

For an initial Fock state, the behavior is similar and the condensed fraction 
remains closer to its initial value for large enough couplings
(Fig.~\ref{fig14}, lower panel). The initial state is in this case the Fock state 
$|21,9\rangle$($c_k =\delta_{k,21}$), with $\langle z_a\rangle (0)=0.4$. 

Finally, in Fig.~\ref{fig15} we present the time correlation for both 
the squeezed and Fock states used in Fig.~\ref{fig14}. For the squeezed case, 
the picture is similar to the case of an initial coherent preparation. The 
two subsystems get correlated for $\Lambda_{ab}\geq 0.01$. In the Fock regime, 
the picture clearly degrades, and although a certain synchronization is found, it 
does not abide in time. 

\section{Conclusion}
\label{sec:conc}

We have considered the coupled dynamics of two ultracold atomic 
clouds, one of which is assumed to be Bose-Einstein condensed during the 
evolution. Our main finding is that by increasing the coupling between 
the two subsystems two net effects take place: (1) the dephasing 
produced by atom-atom interactions in the non condensed subsystem is found 
to decrease as the coupling is increased, and (2) the coherent oscillations of 
both subsystems are found to synchronize. 
This phase-locking phenomenon is characterized by studying the evolution of 
the average population imbalance of each subsystem under different conditions. 
When synchronization appears, the $a$ state is prevented to fragment and 
remains Bose-Einstein condensed: this allows a comparison of the reported 
phase-locking with MS within a classical description.
Even if
the role of the condensate is dominant, preventing the ultracold cloud from losing coherence,
the reported synchronization differs from entrainment.
The ultracold cloud is indeed $driven$ by the
condensate but the latter is also influenced by the feedback of the cloud and both systems evolve towards a different oscillatory dynamics. 
Synchronization
is therefore $ hybrid$ (between a condensate driving the ultracold cloud to remain coherent)
but $mutual$, being the dynamics of both clouds
determined by the reciprocal coupling and in spite of their different regime.

Our results are of relevance for future applications of bimodal 
quantum many-body systems. In particular, since the dephasing arising 
from the atom-atom interactions is found to disappear for large enough 
coupling, we have a way to prevent quantum many-body systems from 
dephasing. Thus, the relevant properties stored in the system, such 
as a large squeezing parameter or a large degree of condensation, are 
preserved during the time evolution if the system is coupled 
to a condensed one. The hybrid synchronization described, which 
appears together with the coherent evolution, can be used as an observable 
control parameter for the phase coherent evolution.

\begin{acknowledgments}
This work was supported by China Scholarship Council, the National Natural Science
Foundation of China (No. 11104217, No. 11205121, and No. 11402199). We acknowledge also partial
financial support from the DGI (Spain) Grant No.FIS2011-24154, FIS2014-54672-P, and FIS2014-60343-P,
the Generalitat de Catalunya Grant No. 2014SGR-401, and EU project 
QuProCS (Grant Agreement 641277). B. J-D. is
supported by the Ram\'on y Cajal program.
\end{acknowledgments}

\end{document}